%% file: main.tex
\def\fulll{}
\pdfoutput=1

\documentclass{article}
\usepackage{fullpage}

\input{preamble}
\usepackage{subfigure}
\iftrue\fi

\title{Distributed Balanced Partitioning via Linear
  Embedding\ifabstract\thanks{Certain
    additional discussions are left out due to space constraints. Refer to the full
    version~\cite{ABM-full} for details.}\else\thanks{A shorter
  version of this work appears in WSDM~2016.}\fi}
\ifx\fulll\undefined
\numberofauthors{3}
\author{
\alignauthor
Kevin Aydin\\
Google Research\\
76 Ninth Ave\\
New York, NY 10011\\
\email{kaydin@google.com}
\alignauthor
MohammadHossein~Bateni\\
Google Research\\
76 Ninth Ave\\
New York, NY 10011\\
\email{bateni@google.com}\\
\alignauthor
Vahab Mirrokni\\
Google Research\\
76 Ninth Ave\\
New York, NY 10011\\
\email{mirrokni@google.com}
}
\else
\author{\centerline{\hfil Kevin Aydin \hfil MohammadHossein Bateni \hfil Vahab  Mirrokni\hfil}\\Google Research\\76 Ninth Avenue, New York, NY 10011\\\texttt{\{kaydin, bateni, mirrokni\}@google.com}}
\date{}
\fi

\begin{document}
\ifabstract
\CopyrightYear{2016} 
\setcopyright{rightsretained} 
\conferenceinfo{WSDM 2016}{February 22-25, 2016, San Francisco, CA, USA} 
\isbn{978-1-4503-3716-8/16/02}
\doi{http://dx.doi.org/10.1145/2835776.2835829}
\fi

\maketitle

\begin{abstract}
\input{abstract}
\end{abstract}


\ifx\fulll\undefined
\terms{Algorithms, Experimentation, Performance}
\fi

\section{Introduction}\label{sec:intro}
\input{intro}

\section{Preliminaries}\label{sec:defs}
\input{defs}
\subsection{Our algorithm}\label{sec:algo}
\input{algo-main}

\section{Initial mapping to line}
\subsection{Random mapping}
\input{random}
\subsection{Hilbert curve mapping}\label{sec:hilbert}
\input{hilbert}
\subsection{Affinity-based mapping}\label{sec:aff-cluster}
\input{affinity}

\section{Improve ordering using semi\-local moves}
\input{metricswap}

\section{Imbalance-inducing postprocessing}\label{sec:post}
\input{postprocess}
\subsection{Dynamic program to optimize cuts}\label{sec:dp}
\input{dp}
\subsection{Scalable linear boundary optimization}\label{sec:lin-opt}
\input{linear-opt}
\subsection{Scalable minimum-cut optimization}
\input{mincut-opt}

\section{Empirical studies}\label{sec:empirical}
\input{practice-main}
\subsection{Datasets}
\input{datasets}
\subsection{Comparison of different Techniques}
\input{ingreds}
\subsection{Comparison to Previous Work}
\input{comparison-facebook}

\input{comparison-fennel}

\input{comparison-friendster}
\subsection{Scalability}\label{sec:scale}
\input{scale}
\subsection{Google Maps Driving Directions}\label{sec:apps}
\input{apps}

\ifabstract
\vspace{-0.4cm}
\fi

\section{Conclusion}
\input{conclusion}

\ifabstract
\vspace{-0.4cm}
\fi

\iffullversion
\bibliographystyle{alpha}
\bibliography{main-long}
\else
\bibliographystyle{acm}
\setlength{\bibsep}{1pt}
\renewcommand{\bibfont}{\small}
\bibliography{main}
\fi

\end{document}

%% file: preamble.tex
\newif\ifabstract
\abstracttrue
\ifdefined\fulll
\abstractfalse
\fi
\newif\iffullversion
\ifabstract \fullversionfalse \else \fullversiontrue \fi
\usepackage{todonotes}
\usepackage{xspace}
\usepackage{amssymb,amsmath}
\iffullversion
\usepackage{amsthm}
\fi
\usepackage{enumitem}

\usepackage{subfigure}
\usepackage{tikz}

\usepackage{color}
\ifabstract\usepackage[numbers]{natbib}\fi
\usepackage
   [breaklinks,bookmarks,bookmarksnumbered,bookmarksopen,bookmarksopenlevel=2]
   {hyperref}
 \definecolor{orange}{rgb}{1,0.5,0}
 \definecolor{darkblue}{rgb}{0,0,0.8}
 \hypersetup{
     colorlinks,
     citecolor=darkblue,
     filecolor=black,
     linkcolor=magenta,
     urlcolor=black
 }
 {\makeatletter \hypersetup{pdftitle={\@title}}}

\ifabstract

\fi

\usepackage[noend]{algorithmic}
\usepackage{algorithm}
\newcommand{\GETS}{\ensuremath{\leftarrow}}

\newtheorem{theorem}{Theorem}
\newtheorem{lemma}[theorem]{Lemma}

\newcommand{\Real}{\ensuremath{\mathbb{R}}\xspace}
\newcommand{\set}[1]{\ensuremath{\{#1\}}\xspace}

\newcommand{\worldg}{\texttt{World-roads}\xspace}
\newcommand{\twitterg}{\texttt{Twitter}\xspace}
\newcommand{\liveg}{\texttt{LiveJournal}\xspace}
\newcommand{\friendg}{\texttt{Friendster}\xspace}

\newcommand{\affinity}{\texttt{AffinityOrdering}\xspace}
\newcommand{\randinit}{\texttt{RandInit}\xspace}
\newcommand{\affdist}{\texttt{AffInvDist}\xspace}
\newcommand{\affcomm}{\texttt{AffCommNeigh}\xspace}
\newcommand{\hilbert}{\texttt{Hilbert}\xspace}

\newcommand{\affmetric}{\texttt{Aff+\-Metric}\xspace}
\newcommand{\affswap}{\texttt{Aff+\-Swap}\xspace}

\newcommand{\affmincut}{\texttt{Aff+\-MinCut}\xspace}

\newcommand{\algmetric}{\texttt{Metric}\xspace}
\newcommand{\algswap}{\texttt{Swap}\xspace}
\newcommand{\algswapfull}{\texttt{RankSwap}\xspace}
\newcommand{\champion}{\texttt{Combination}\xspace}

\newcommand{\mincut}{\texttt{MinCut}\xspace}
\newcommand{\linopt}{\texttt{LinOpt}\xspace}

\newcommand{\mathC}{\ensuremath{\mathcal{C}}\xspace}
\newcommand{\inv}[1]{\ensuremath{#1^{-1}}\xspace}

\ifabstract
\setlist{nolistsep}
\fi

%% file: abstract.tex
Balanced partitioning is often a crucial first step in solving
large-scale graph optimization problems\iffullversion, e.g., \else:
\fi in some cases, a big graph \iffullversion can be \else is \fi
chopped into pieces that fit on one machine to be processed
independently before stitching the results together, leading to
certain suboptimality from the interaction among different pieces. In
other cases, links between different parts may show up in the running
time and/or network communications cost, hence the desire to have
small cut size.

We study a distributed balanced partitioning problem where the goal is
to partition the vertices of a given graph into $k$
pieces\iffullversion\ so as to minimize \else, minimizing \fi the
total cut size. Our algorithm is composed of a few steps that are
easily implementable in distributed computation frameworks, e.g.,
MapReduce.  The algorithm first embeds nodes of the graph onto a line,
and then processes nodes in a distributed manner guided by the linear
embedding order. We examine various ways to find the first embedding,
e.g., via a hierarchical clustering or Hilbert curves. Then we apply
four different techniques \iffullversion including \else such as \fi
local swaps, minimum cuts on \iffullversion the boundaries of
partitions\else partition boundaries\fi, as well as contraction and
dynamic programming.

\iffullversion As our empirical study, we compare 
\else Our empirical study compares \fi
the above techniques with each
other, and \iffullversion also \fi to previous work in distributed
\iffullversion graph \fi algorithms,
e.g., a label propagation method~\cite{wsdm-paper},
FENNEL~\cite{fennel14} and Spinner~\cite{Spinner14}.  We report our
results both on a private map graph and several public social
networks, and show that our results beat previous distributed
algorithms: \iffullversion e.g., compared to the label propagation
algorithm~\cite{wsdm-paper}, we report an improvement of $15$-$25\%$
in the cut value\else we notice, e.g., $15$-$25\%$ reduction in cut size
over \cite{wsdm-paper}\fi. We also observe that our algorithms allow for
scalable distributed implementation for any number of partitions.
Finally, we apply our techniques for the Google Maps Driving
Directions to minimize the number of multi-shard queries with the goal
of saving in CPU usage.  During live experiments, we observe an
$\approx 40\%$ drop in the number of multi-shard queries when
comparing our method with a standard geography-based method.

%% file: intro.tex
Graph partitioning is a crucial first step in developing a tool for
mining big graphs or solving large-scale optimization problems. In
many applications, the partition sizes have to be balanced or almost
balanced so that each part can be handled by a single machine to
ensure the speedup of parallel computing over different parts. Such a
balanced graph partitioning tool is applicable throughout scientific
computation for dealing with distributed computations over massive
data sets, and serves as a tradeoff between \iffullversion the \fi
local computation and total communication amongst machines. In several
applications, a graph serves as a substitute for the computational
domain~\cite{dongarra2003-parallel}. Each vertex denotes a piece of
information and edges denote \iffullversion dependencies between
information. \else their dependencies. \fi Usually, the goal is to
partition the vertices where no part is too large, and the number of
edges across parts is small\iffullversion. Such a partition implies
\else, implying \fi that most of the work is within a part and only
minimal work takes place between parts. In these applications, links
between different parts may show up in the running time and/or network
communications cost. Thus, a good partitioning minimizes the amount of
communication during a distributed computation. In other applications,
a big graph \iffullversion can be \else is \fi carefully chopped into
parts that fit on one machine to be processed independently before
finally stitching the individual results together, leading to certain
suboptimality arising from the interaction between different parts. To
formally capture these situations, we study a balanced partitioning
problem where the goal is to partition the vertices of a given graph
into $k$ parts so as to minimize the total cut size.\footnote{We
  emphasize that our main focus is on the size of the resulting cut
  and not the resource consumption of our algorithm, although it is
  fully distributable, uses linear space and runs for all the relevant
  experiments in reasonable amount of time (more or less comparable to
  the state of the art). Unfortunately, we cannot reveal the exact
  running times due to corporate restrictions. Nevertheless, we report
  in Section~\ref{sec:scale} relative running times of our algorithm
  on synthetic data of varying sizes to prove its scalability.}

This is a challenging problem that is computationally hard even for
medium-size graphs~\cite{AR06} as it captures the graph
bisection~\cite{GJ79}, hence all attempts at tackling it necessarily
relies on heuristics. While the topic of large-scale balanced graph
partitioning has attracted significant attention in the
literature~\cite{DGRW11,DGRW12,fennel14,wsdm-paper}, the large body of
previous work study large-scale but non-distributed solutions to this
problem. Several motivations necessitate the quest for a distributed
algorithm for these problems: (i) first of all, huge graphs with
hundreds of billions of edges that do not fit in memory are becoming
increasingly common~\cite{fennel14,wsdm-paper}; (ii) in many
distributed graph processing frameworks such as
Pregel~\cite{Malewicz2010:Pregel}, Apache
Giraph~\cite{Ching2010:Giraph}, and PEGASUS~\cite{pegasus}, we need to
partition the graph into different pieces, since the underlying graph
does not fit on a single machine, and therefore for the same reason,
we need to partition the graph in a distributed manner; (iii) the
ability to run a distributed algorithm using a common \iffullversion
distributed computing \fi framework such as MapReduce on a
\iffullversion distributed computing \fi platform consisting of
commodity hardware makes an algorithm much more widely applicable in
practice; and (iv) even if the graph fits in memory of a
super-computer, implementing some algorithms requires superlinear
memory which is not feasible on a single machine. The need for
distributed algorithms has been observed by several practical and
theoretical research papers~\cite{wsdm-paper,fennel14}.

For streaming and some distributed optimization models,
Stanton~\cite{S14} has recently shown that achieving formal
approximation guarantees is information theoretically impossible, say, 
for a class of random graphs. Given the hardness of this problem, we
explore several distributed heuristic algorithms. Our
algorithm is composed of a few logical, simple steps, which can be
implemented in various distributed computation frameworks, e.g.,
MapReduce~\cite{osdi-DG04}.

{\bf \noindent An application in Google Maps driving directions.} As
one \iffullversion of the applications \else application \fi of our
balanced partitioning algorithm, we \iffullversion study a balanced
partitioning problem in \else turn to \fi the Google Maps Driving
Directions.  This system computes the optimal driving route for any
given {\em source-destination query} (pairs of points on
\iffullversion Google Maps\else the map\fi). \iffullversion In order
to answer such pairwise queries in this large-scale graph application,
a \else A \fi plausible approach
is to obtain a suitable partitioning of the graph beforehand so that
each server handles its own section of the graph. \iffullversion In this application,
ideally \else Ideally \fi we seek to minimize the number of cross-shard queries,
\iffullversion for which the source and destination are not in the
same shard, i.e., the total cross-shard traffic is as low as possible,
\fi since handling \iffullversion cross-shard queries \else these \fi 
is much more costly and entails extra communication and coordination
across multiple servers. \iffullversion In order to \else To \fi
reduce the cross-shard query traffic we \iffullversion want \else aim
\fi to minimize the cut size between \iffullversion the \fi partitions, \iffullversion
with the expectation \else expecting \fi that less directions will be
produced \iffullversion where  the source and the destination are \else
with source and destination \fi in different
partitions.  Reducing the cut size in practice results in denser
regions in \iffullversion the same \else each \fi partition, therefore
most queries will be contained within one shard. We confirm this
observation with extensive studies using historical query data and
live experiments (\iffullversion see Section~\ref{sec:apps} for more
\fi details\ifabstract\ in \ref{sec:apps}\fi).
Conveniently, a linear embedding provides a very simple way of
specifying partitions by merely providing two numbers for each shard's
boundaries.  Therefore it is straightforward and fast to identify the
shard a given query point belongs to.  As \iffullversion we will
discuss \else discussed \fi later, our
study of linear embedding of graphs while minimizing the cut size is
partly motivated by these reasons.

{\bf \noindent Our Contributions.} 
Our contributions in this paper can be divided into five categories.

1) We introduce a multi-stage distributed optimization framework for
balanced partitioning based on first embedding the graph into a line
and then optimizing the clusters guided by the order of nodes on the
line.\footnote{Embedding into a line is also essential in graph
  compression; see, e.g., \cite{BRSV11,CKLMPR09}.} This framework is
not only suitable for distributed partitioning of a large-scale graph,
but also it is directly applicable to balanced partitioning
applications like those in Google Maps Driving Directions described
above.  \iffullversion More specifically, we develop a general
technique that first embeds nodes of the graph onto a line, and then
post-processes nodes in a distributed manner guided by the order of
our linear embedding.  \fi

2) As for the initialization stage, we examine several ways to find
the first embedding\footnote{It might be worth studying spectral
  embedding methods or standard embedding techniques into $\ell_1$,
  but we did not try them due to not having a scalable distributed
  implementation, and also since the hierarchical clustering-based
  embedding worked pretty well in practice.  We leave this for future
  research.}: (i) by na\"{\i}vely using a random ordering, (ii) for
map graphs by the Hilbert-curve embedding, and (iii) for general
graphs by applying hierarchical clustering on an edge-weighted graph
where the edge weights are based either on the number of common
neighbors of the nodes in the graph, or on the inverse of the distance
of the \iffullversion corresponding \fi nodes on the map. Whereas the
Hilbert-curve embedding and our hierarchical clustering based on node
distances may only be applied to maps graphs, the hierarchical
clustering based on the number of common neighbors is applicable to
all graphs. While all these methods prove useful, to our surprise, the
latter (most general) initial embedding technique is the best
initialization technique even for maps.  We later explain that all
these methods have very efficient and scalable distributed
implementations.  \iffullversion See
the discussion in Section~\ref{sec:aff-cluster} and
Footnote~\ref{footnote:cc} regarding the efficiency of the
hierarchical clustering, and the references in
Footnote~\ref{footnote:triangle} regarding the challenge to construct
the similarity metric.  \fi

3) As a next step, we apply four methods to postprocess the initial
ordering and produce an improved cut. The methods include a metric
ordering optimization method, a local improvement method based on
random swaps, another based on \iffullversion computing \fi minimum
cuts in the \iffullversion boundaries of partitions\else partition
boundaries\fi, and finally a technique based on contracting the
min-cut-based clusters, and applying dynamic programming to compute
optimal cluster boundaries. Our final algorithm (called \champion)
combines the best of our initialization methods, and iterates on the
various postprocessing methods until it converges.  The resulting
algorithm is quite scalable: it runs smoothly on
graphs
with hundreds of millions of nodes and billions of edges.

4) As our empirical study, we compare the above techniques with each
other, and more importantly with previous work. In particular, we
compare our results to recent work in (distributed) balanced graph
partitioning, including a recent label propagation-based
algorithm~\cite{wsdm-paper}, FENNEL~\cite{fennel14},
Spinner~\cite{Spinner14} and
METIS~\cite{karypis1998-metis}.\footnote{Some works \iffullversion such as
  \cite{SK12} \fi are indirectly compared to, and others
  \iffullversion such as \cite{Blogel}) \else (e.g., \cite{Blogel}) \fi did not report cut sizes. We relied for these
  comaprisons on available cut results for a host of large public
  graphs.}  We report our results on both a large private map graph,
and also on several public social networks studied in previous
work~\cite{wsdm-paper,fennel14,Spinner14,Blogel}.
\iffullversion\begin{itemize}[topsep=0pt,parsep=0pt,partopsep=0pt] 
\item \fi First, we show that our distributed algorithm consistently
  beats the label propagation algorithm by Ugander and
  Backstrom~\cite{wsdm-paper} (on \liveg) by a reasonable margin for
  all values of $k$, e.g., for $k=20$, we improve the cut by $25\%$
  (from $37\%$ to $27.5\%$ of total edge weight), and for $k=100$, we
  improve the cut by $15\%$ (from $49\%$ to $41.5\%$).  The clustering
  outputs can be found in~\cite{public-output}.
  \iffullversion\item\fi In addition, for $k>2$ partitions, we show
  that our algorithm beats METIS and FENNEL (on \twitterg) by a
  reasonable factor.  For $k=2$, the results that we obtain beats the
  output of METIS, but it is slightly worse than the result reported
  by FENNEL.  \iffullversion\item\fi For both graphs, our results are
  consistently superior to that of Spinner~\cite{Spinner14} with a
  wide margin.
\iffullversion\item\fi
We also note that our algorithms allow for scalable
  distributed implementation for small or large number of
  partitions. More specifically, changing $k$ from $2$ to tens of
  thousands does not change the running time significantly.  
\iffullversion\item\fi
As for comparing various initialization methods, we observe that while
  geographic Hilbert-curve techniques outperform the random ordering,
  the methods based on the hierarchical clustering using the number of
  common neighbors as the similarity measure between
  nodes\footnote{This is needed only once to obtain the initial graph,
    and the new edge weights in the subsequent rounds are computed via
    aggregation. Further note that approximate triangle counting can
    be done efficiently in MapReduce; see, e.g., \cite{TKM2009}.\label{footnote:triangle}}
  outperform the geography-based initial embeddings, even for map
  graphs.
\iffullversion\item\fi
As for comparing other techniques, we observe the random
  swap techniques are effective on \twitterg, and the min-cut-based or
  contract and dynamic program techniques are very effective for the
  map graphs. Overall, we realize that these techniques complement
  each other, and combining them in a \champion algorithm is the most
  effective method.
\iffullversion\end{itemize}\fi

5) As mentioned earlier, we apply our results to the Google Maps
Driving Directions application, and deploy two linear-embedding-based
algorithms on the World's Map graph. We first examine the best
imbalance factor for our cut-optimization technique, and observe that
we can reduce $21\%$ of cross-shard queries by increasing the
imbalance factor from $0\%$ to $10\%$.  The two methods that we
examined via live experiments were (i) a baseline approach based on
the Hilbert-curve embedding, and (ii) one method based on applying our
cut-optimization post-processing techniques. By running live
experiments on the real traffic, we observe the number of multi-shard
queries from our cut-optimization techniques is 40$\%$ less compared
to the baseline Hilbert embedding technique. This, in turn, results in
less CPU usage in response to queries\footnote{The exact decrease in
  CPU usage depends on the underlying serving infrastructure which is
  not our focus and is not revealed due to company policies. We note
  that there are several other algorithmic techniques and system
  tricks that are involved in setting up the distributed serving
  infrastructure for Google Maps driving directions. This system is
  handled by the Maps engineering team, and is not discussed here as
  it is not the focus of this paper.}.

{\bf\noindent Other Related Work.}
Balanced partitioning is a challenging problem to  approximate within a constant
factor~\cite{GoldschmidtH88,AR06,FK02}, or to approximate
within any factor in several distributed or steaming models~\cite{S14}.
As for heuristic algorithms for this problem in the distributed or
streaming models, a number of recent papers have been written about
this topic~\cite{SK12,wsdm-paper,fennel14}. Our algorithms are
different from the ones studied in these papers. The most similar
related work are the label propagation-based methods of Ugander and
Backstrom~\cite{wsdm-paper} and Martella et al.\
(Spinner)~\cite{Spinner14} which develop a scalable distributed
algorithm for balanced partitioning.  Our random swap technique is
similar in spirit to the label propagation algorithm studied by
\cite{wsdm-paper}, however, we also examine three other methods as a
postprocessing stage and find out that these methods work well in
combination.  Moreover, \cite{wsdm-paper} studied two different
methods for their initialization, a random initialization, and a
geographic initialization. We also examine random ordering, and a
Hilbert-curve ordering which is similar to the geographic
initialization by \cite{wsdm-paper}, however we examine two other
initialization techniques and observe that even for map-based
geographic graphs, the initialization methods based on hierarchical
clustering outperform the geography-based initial ordering.  Overall,
we compare our algorithm directly on a \liveg public graph (the only
public graph reported in~\cite{wsdm-paper}), and improve the cut
values achieved in~\cite{wsdm-paper,Spinner14} by a large margin for
all values of $k$.  In addition, algorithms developed in
\cite{SK12,fennel14} are suitable for the streaming model, but one can
implement variants of those algorithms in a distributed manner. We
also compare our algorithm directly to the numbers reported on the
large-scale \twitterg graph by FENNEL~\cite{fennel14}\iffullversion, and show that
our algorithm compares favorably with the FENNEL output, hence
indirectly comparing against \cite{SK12} because of the comparison
provided in \cite{fennel14}\else, demonstrating the performance of our algorithm\fi.


\iffullversion Motivated by a variety of big data applications,
distributed clustering has attracted significant attention over the
literature, both for applications with no strict size constraints on
the clusters~\cite{alina-kdd-paper,ravi-kmeans-paper,nips-BEL13}, or
with explicit or implicit size constraints~\cite{BBLM14}.  A main
difference between such balanced clustering problems and the balanced
graph partitioning problems considered here is that in the graph
partitioning problems a main objective function is to minimize the cut
function, but in those clustering problems, the main goal is minimize
the maximum or average distance of nodes to the center of those
clusters.  \fi

\ifabstract
The reader is referred to the full version of the
paper~\cite{public-output} to see the omitted proofs and discussions
as well as additional related work.
\fi

%% file: defs.tex
For an integer $n$, we define $[n] = \set{1, 2, \dots, n}$. We also
slightly abuse the notation and, for a function $f : A \mapsto \Real$,
define $f(A') = \sum_{a\in A'} f(a)$ if $A' \subseteq A$.
For a function $f: S \mapsto T$, we let $\inv f(t)$ for $t\in T$ 
denote the set of elements in $S$ mapped to $t$ via $f$. 
%
%
For a permutation $\pi$, let $\pi_i$ denote, for $i \in [n]$, the
element in position $i$ of $\pi$. Moreover, let $\pi(i \rightarrow j) =
\set{\pi_i, \pi_{i+1}, \dots, \pi_j}$ for $i \in [n], j \in [n] \cup
\set{0}$. Note that, in particular, $\pi(i \rightarrow j ) = \emptyset$ if $j <
i$.

\subsection{Problem definition}
Given is a graph $G(V, E)$ of $n$ vertices with edge lengths and node
weights, as well as an integer $k$, and a real number $\alpha \geq 0$.
Let us denote the edge lengths by $d : E \mapsto \Real$, and the
vertex weights by $w : V \mapsto \Real$.  
%
A partition of vertices of $G$ into $k$ parts $\set{V_i: i\in [k]}$ is
said to be $\alpha$-balanced if and only if
\iffullversion $$(1-\alpha)\frac{w(V)}{k} \leq w(V_i) \leq (1+\alpha)
\frac{w(V)}{k}.$$ \else $(1-\alpha) w(V) / k \leq w(V_i) \leq
(1+\alpha) w(V) / k$. \fi In particular, a zero-balanced (or fully
balanced) partition is one where all partitions have the same weight.
The cut length of the partition is the total sum of all edges whose
endpoints fall in different parts:
\iffullversion
\begin{align*}
\sum_i\sum_{j<i} \sum_{u\in V_i}\sum_{v\in V_j: (u,v) \in E(G)} w(u,v).
\end{align*}

\fi
Our goal is to find an $\alpha$-balanced partition whose cut size is
(approximately) minimized.

\ifabstract The problem is NP-hard (the case of $\alpha = 0, k = 2$ is
the minimum bisection problem).  For arbitrary $k,
\alpha$, we get the minimum balanced $k$-cut problem, that is known to
be inapproximable within any finite factor~\cite{AR06}; the best
approximation factor for it is $O(\log^{1.5}n)$ (if $\alpha > 0$ is
constant).  
\else
The problem is NP-hard as the case of $\alpha = 0, k
= 2$ is equivalent to the minimum bisection
problem~\cite{gareyjohnson}.  For arbitrary $k$, we get the minimum
balanced $k$-cut problem, that is known to be inapproximable within
any finite factor~\cite{AR06}; the best known approximation factor for it is
$O(\log^{1.5}n)$ (if $\alpha > 0$ is constant).  \fi


%% file: algo-main.tex
Our algorithm consists of three main parts:
\iffullversion 
\begin{enumerate}
\item We first find a suitable mapping of the vertices to a line. This
  gives us an ordering of the vertices that presumably places (most)
  neighbors close to each other, therefore somewhat reduces the
  minimum-cut partitioning problem to an almost local optimization
  one.
\item We next attempt to improve the ordering mainly by swapping
  vertices in a semilocal manner. These moves are done so as to
  improve certain metrics (perhaps, the cut size of a fully balanced
  partition).
\item Finally, we use local postprocessing optimization in the ``split
  windows'' (i.e., a small interval around the equal-size partitions
  cut points taking into account permissible imbalance) to improve the
  partition's cut size.
\end{enumerate}
\fi
\ifabstract 
(i) We first find a suitable mapping of the vertices to a line. This
gives us an ordering of the vertices that presumably places (most)
neighbors close to each other, therefore somewhat reduces the
  minimum-cut partitioning problem to an almost local optimization;
(ii) then, we attempt to improve the ordering mainly by swapping
  vertices in a semilocal manner. These moves are done so as to
  improve certain metrics (perhaps, the cut size of a fully balanced
  partition), and
finally, (iii) we use local postprocessing optimization in the ``split
  windows'' (i.e., a small interval around the equal-size partitions
  cut points taking into account permissible imbalance) to improve the
  partition's cut size.
%
\fi

\begin{figure}
\ifabstract\centerline{\includegraphics[width=0.6\columnwidth]{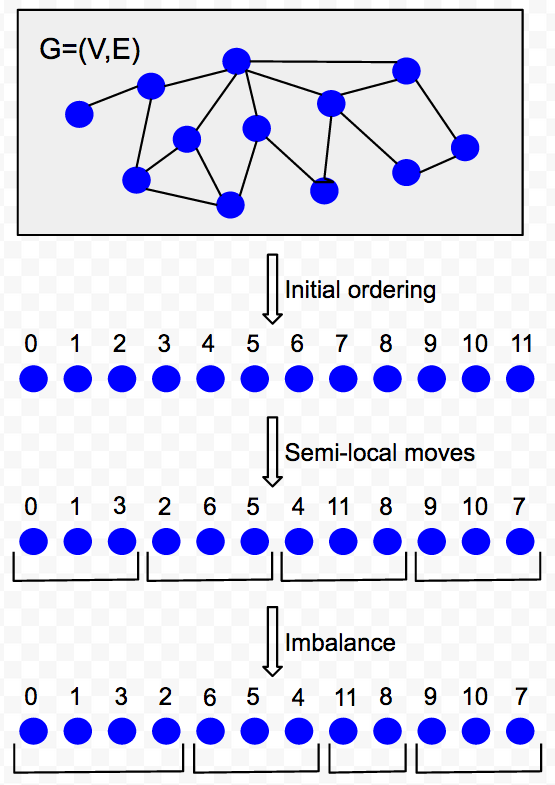}}\fi
\iffullversion%
\centerline{\includegraphics[width=0.6\columnwidth]{img/alg-outline.png}}%
\fi
\vspace{-1mm}
\caption{A natural flow for our algorithm. The vertices of $G$ are first
  embedded on a line, then the embedding is improved via swaps, and
  finally imbalanced clusters are created according to the
  ordering of vertices. As we explain later on, iterating on these
  steps yields better results.\label{fig:alg-main}}
\end{figure}

Note that one implementation may use only some of the above, and
clearly any implementation will pick one method to perform each task
above and mix them to get a final almost balanced partition.  Indeed,
we report some of our results (specially those in comparison to the
previous work) based on \champion, which uses \affcomm to get the
initial ordering and then iteratively applies techniques from the
second and third stages above (e.g., \algmetric, \algswap, \mincut)
until the result converges.  (The convergence in our experiments
happens in two/three rounds.)

\begin{algorithm}
\caption{$\champion(G,k,\alpha)$\label{alg:combination}}
\textbf{Input:} Graph $G$, number of parts $k$, imbalance
param.\ $\alpha$\\
\textbf{Output:} A partition of $V(G)$ into $k$ parts

\begin{algorithmic}[1]
  \FORALL {$(u, v) \in E(G)$}
    \STATE $w(u, v) \GETS $ number of common neighbors of $u, v$
  \ENDFOR
  \STATE $\pi\GETS\affinity(G,w)$
  \FOR {$i = 0$ \TO $k$}
    \STATE $q(i) \GETS \lfloor\frac{in}{k}\rfloor$ \COMMENT{fully
      balanced split points}
  \ENDFOR
  \REPEAT
  \STATE $(\pi', q') \GETS (\pi, q)$
  \STATE Use any (semilocal or
  imbalance-inducing) postprocessing technique on $(G, \pi', q')$ to obtain $(\pi, q)$
  \UNTIL {$\pi = \pi'$ \AND $q = q'$}
  \FORALL {$i \in [k]$}
    \STATE $V_i \GETS \pi(q(i)+1 \rightarrow q(i+1))$ 
  \ENDFOR
  \RETURN $\set{V_1, V_2, \dots, V_k}$
\end{algorithmic}
\end{algorithm}

%% file: random.tex
The easiest method to produce an ordering for the vertices is to randomly permute them.  Though very fast, this method does not seem to lead to much progress towards our goal of finding good cuts.  In particular, if we then turn this ordering into a partition by cutting contiguous pieces of equal size, we end up with a random partitiong of the input (into equal parts) which is almost surely a bad cut: a standard probablistic argument shows that the cut has expected ratio $1-\frac{1}{k}$.
  Nevertheless, the next stage of the algorithm (i.e., the semilocal optimization by swapping) can generate a relatively good ordering with this na\"{i}ve starting point.


%% file: hilbert.tex
For certain graphs, geographic/geometric information is available for
each vertex.  It is fairly easy, then, to construct an ordering using
a space-filling curve---prime examples are Peano, Morton and Hilbert
curves, but we focus on the latter in this work.  These methods are
known to capture proximity well: nodes that are close in the space are
expected to be placed nearby on the line.

There has been extensive study of the applicability of these methods
in solving largescale optimization problems in parallel; see, e.g.,
\cite{book:global-opt-parallel,article:MJFS01:analysis-hilbert}.

Not only can this algorithm be used on its own without any cut
guarantees, but it can also be employed to break a big instance down
into smaller ones that we can afford to run more intensive
computations on.  Both applications were known previously.

The previous works do not offer any theoretical guarantees on the
quality of the cut generated from Hilbert curves.  However, certain
assumptions on the distributions of edge lengths and node positions
let us bound the resulting cut ratio, and show that it is
significantly less than the result of random
ordering\iftrue~\cite{GL96,NRS02,FTW00}\fi.  This is observed
in our experiments, too.



%% file: affinity.tex
One drawback of the Hilbert curve cover (even when coordinates are
available) is that it ignores the actual edges in the graph.  For an
illustration, consider using the Hilbert curve to map certain points
in an archipalego (or just a small number of islands or peninsulas).
The Hilbert curve, unaware of the connectivities, traverses the
vertices in a semirandom order, therefore, it may jump from island to
island without covering the entire island first.

To address this issue, we use an agglomerative hierarchical clustering
method, usually called average-linkage clustering\footnote{Typically
  attributed to \cite{SM58}, this is reminiscent of a parallel version
  of {Bor\.uvka}'s algorithm for minimum spanning tree where the
  $\min$ operation is replaced by average\iffalse~\cite{Boruvka}\fi.}.
We call the method ``affinity clustering'' since it takes into account
the affinity of vertices.  Informally, every node starts in a
singleton cluster of its own, and then in several stages we group
vertices that are closely connected, hence building a tree of these
connections.  More precisely, at every stage, each node turns on the
connection to its closest neighbor, after which the connected
components form clusters.  The similarities used in the next level
between constructed clusters are computed via some function of the
similarities between the elements forming the clusters; in particular,
we take the average function for this purpose.

The final ordering is produced by sorting the vertex labels produced
as follows.  Let the label for each vertex be the concatenation of
vertex ID strings from the root to the corresponding leaf.  
Sorting the constructed labels places the vertices under each branch
in a contiguous piece on the line.  The same guarantee holds
recursively, hence the (edge) proximity is preserved well.  A
pseudocode for this procedure is given in
Algorithm~\ref{alg:affinity}: Line~\ref{alg:line:concat} uses notation
$a\#b$ to denote the concatenation of labels for $a$ and $b$.  Notice
that this pseudocode explains a sequential algorithm, however, the
procedure can be efficiently implemented in a few rounds of MapReduce.
In fact, our algorithm is very similar to \cite{RastogiMCS13} but uses
Distributed Hash Tables as explained in
\cite{KiverisLMRV14}.\footnote{In fact, as mentioned therein, an
  implementation of the connected components algorithm within the same
  framework and using the same techniques provides a 20-40 times
  improvement in running time compared to best previously known
  algorithms.\label{footnote:cc}}

\begin{algorithm}
\caption{$\affinity(G,w)$\label{alg:affinity}}
\textbf{Input:} Graph $G$, (partial) similarity function $w$\\
\textbf{Output:} A permutation of vertices

\begin{algorithmic}[1]
  \STATE $\mathC^0$ \GETS \set{ \set{v} | v \in V(G)}
  \FOR {$v \in V(G)$}
    \STATE $\ell(v) \GETS v$
  \ENDFOR
  \STATE $i \GETS 0$
  \REPEAT  
    \FOR {$s \in \mathC^i$}
      \STATE $\displaystyle p(s) \GETS \arg\max_{t \in C^i} w(s, t)$
    \ENDFOR
    \STATE $\mathC^{i+1} \GETS \emptyset$
    \FOR {$t \in \set{p(s) | s \in \mathC^i }$}
      \STATE $S \GETS \inv{p}(t)$
      \STATE $S' \GETS \bigcup_{s \in S}s$
      \STATE $\displaystyle \mathC^{i+1} \GETS \mathC^{i+1} \cup S'$
      \STATE $\displaystyle r \GETS \arg\min_{r' \in S'} s$
      \FOR {$s' \in S'$}
        \STATE $\ell(s') \GETS r\#\ell(s')$ \label{alg:line:concat}
      \ENDFOR
    \ENDFOR
    \STATE $i \GETS i + 1$
  \UNTIL {$|\mathC^i| = |\mathC^{i-1}|$}
  \STATE $\pi \GETS$ permutation of vertices sorted according to
  $\ell$
  \RETURN $\pi$
\end{algorithmic}
\end{algorithm}

\begin{figure}
\ifabstract
\centerline{\includegraphics[width=.9\columnwidth]{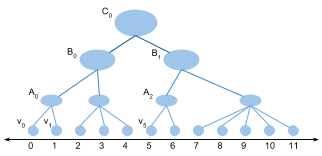}}
\else
\centerline{\includegraphics[width=.6\columnwidth]{img/affinity-tree.png}}
\fi
\vspace{-1mm}
\caption{Illustration of Affinity tree ordering. The nodes at the
  bottom row correspond to the vertices of the graph, and the nodes at
  the higher levels are formed by merging nodes from lower levels.
  The vertices are finally sorted in the picture according to their
  labels.  For instance, the label for $v_0$ is $C_0\#B_0\#A_0\#v_0$
  whereas $v_5$ is labeled $C_0\#B_1\#A_2\#v_5$.\label{fig:affinity}}
\vspace{-3mm}
\end{figure}

Intuitively, we expect vertices connected in lower levels
(farther from the root) to be closer than those connected
in later stages.  (See Figure~\ref{fig:affinity} for illustration.)  In
particular, we observed that a graph with several connected components
gives an advantage to affinity tree ordering over the ordering
produced by Hilbert curve.

For this approach to work, we require meaningful
distances/similarities between vertices of the graph.  Conveniently it
does not matter whether we have access to distances or similarity
values.  However, though theoretically possible to run the affinity
clustering algorithm on an unweighted graph with distances $1$ and
$\infty$ for edges and non-edges respectively, this leads to a lot of
arbitrary tie-breaks that makes the result
irrelevant.\footnote{Practically the worse outcome is a highly
  unbalanced hierarchical partitioning because arbitrary tie-breaks
  favor the ``first'' cluster.}

Fortunately, there are standard ways to impose a metric on an
unweighted graph: e.g., common neighbors ratio or personalized page
rank.  We focus on the former metric and compute for every pair of
neighbors the ratio of the number of their common neighbors over the
total number of their distinct neighbors.  Computing the number of
common neighbors can be implemented using standard techniques in
MapReduce, however, one can use sampling when dealing with
high-degree vertices to improve the running time and memory footprint
(leading to an approximate result).\footnote{As alluded earlier in the
text, there are very fast distributed algorithms for computing this
metric; see, e.g., \cite{TKM2009}.}


%% file: metricswap.tex
Given an initial ordering of vertices for example by \affinity, we can
further improve the cut size by applying semilocal moves. The
motivation here is that these simple techniques provide us with highly
parallelizable algorithms that can be iteratively applied (e.g., by a
MapReduce pipeline) so the result converges to high-quality
partitions. In the following we discuss two such approaches.

\subsection{Minimum Linear Arrangement} \label{minla}

Minimum Linear Arrangement (MinLA) is a well studied NP-hard
optimization problem~\cite{GJ79}.\footnote{In fact, the problem admits
  $O(\log\log n)$ approximation on planar metrics~\cite{RaoR04} and
  $O(\sqrt{\log n}\log\log n)$ on general graph
  metrics~\cite{FeigeL07,CharikarHKR10}.}  Given an
undirected graph $G=(V,E)$ (with edge weights $w_{uv}$) we seek a
one-to-one function $\phi: V \rightarrow \set{0,1,\dots,n-1}$ that
minimizes
\begin{align}\label{eqn:metric}
     \sum_{(u, v) \in E} |\phi(v) - \phi(u)|w_{uv}.
\end{align}

Optimizing the sum above results in a linear ordering of vertices
along a line such that neighbor nodes are nearby each other. The end
result is that dense regions of the graph will be isolated into
clusters and therefore makes it easier to identify cut boundaries on
the line.

We have implemented a very simple distributed algorithm (in
MapReduce) that directly optimizes MinLA. The algorithm simply
iterates on the following MapReduce phases until it converges (or runs
up to a specified number of steps):
\begin{itemize}
\item \textbf{MR Phase 1:} Each node computes its optimal rank as the
  weighted median\footnote{All else fixed, moving a node to the
    weighted median of its neighbors---a standard computation in
    MapReduce---optimizes Equation~\ref{eqn:metric}.} of its neighbors
  and outputs its new rank.
\item \textbf{MR Phase 2:} Assigns final ranks to each node (i.e.,
  handles duplicate resolution, for example, by simple ID-based
  ordering).
\end{itemize}

%
Note that since Phase 1 is done in parallel for each node and
independently of other moves, there will be side effects, hence it is
an optimistic algorithm with the hope that it will converge to a
stable state after few rounds.  Our experimental results show that in
practice this algorithm indeed converges for all the graph types that
we tried, although the number of rounds depends on the underlying
graph.

\subsection{Rank Swap}
Given an existing linear ordering of vertices in a graph we can
further improve the cut size (of the fully balanced partition based on
it) via semilocal swaps. Note that unlike MinLA heuristic
discussed in Section~\ref{minla}, this process depends on the
prechosen cut boundaries, i.e., the number of final partitions $k$.
Notice that one expects that the semilocal swap operations will
be effective once a good initial ordering provided by either the
Affinity-based mapping and/or MinLA.  Indeed our experimental results
show that this procedure is extremely effective and produces cut sizes
better than the competition on some public social graphs we tried.

\algswapfull can be implemented in a straightforward way on a
distributed (for example, MapReduce) framework as outlined by
Algorithm~\ref{alg:rankswap}.
  
\begin{algorithm}
\caption{$\algswapfull(G,k,r)$\label{alg:rankswap}}
\textbf{Input:} Graph $G$, number of partitions $k$, number of intervals per partition $r$\\
\textbf{Output:} A partition of $V(G)$ into $k$ parts

\hspace{5mm}{\bf Controller}:
\begin{algorithmic}[1]
 \STATE $\quad$ Pair nearby partitions
 \STATE $\quad$ Split each partition into $r$ intervals $I_0, \dots, I_{r-1}$ and randomly pair intervals between paired partitions\\
{\bf Map} $\langle I_i; I_j\rangle$:
 \REPEAT
  \STATE Pick node pair $u \in I_i$, $v \in I_j$ with the best cut improvement
  \STATE Swap $u$ and $v$
 \UNTIL no pair improves the cut
 \FORALL {$u \in I_i \cup I_j$}
 \STATE Emit $\langle u; r_u \rangle$. \COMMENT{$r_u$ is the new rank of $u$}
 \ENDFOR 
{\bf Reduce} $\langle u;  r_u \rangle$:
\STATE $\quad$ Emit $\langle u;  r_u \rangle$. \COMMENT{Identity Reducer}
\end{algorithmic}

\end{algorithm}

Subdividing each partition into intervals is important, especially for
small number of partitions $k$, to achieve parallelism, and allowing
possibly more time- and memory-consuming swap operations.  Pairing the
intervals between two partitions can be done in various ways. One
simple example is random pairing.  Our experiments showed that this is
almost as effective as more complicated ones, hence we do not discuss
the other methods in detail here.

\ifabstract The swap operation between two intervals (handled by each
Mapper task) can be done in various ways. We have experimented with
several methods each with increasing complexity and settled on the
following one: 1) Sort the nodes in each interval in descending order
by their cut size reduction.  2) Starting from the top node in the
first interval find a best pair (if one exist) from the second
interval.  3) If a pair is found swap the nodes and update all their
neighbors and iterate.
         
Our extensive studies with this method confirmed that it gives the
best results in terms of quality at the expense of an acceptable
additional cost. It runs fast in practice due to the relatively small
sizes of the intervals because of the parallelism we achieve. We can
also afford an additional cost in the swap method because each
interval pair is handled by a separate Mapper task.
  
\fi 

\iffullversion
The swap operation between two intervals (handled by each Mapper task)
can be done in various ways. (The version described in Algorithm \ref{alg:rankswap} is Method 3 below.)

Each approach below starts with computing the cut size reduction as a result of moving a node from one interval to the other. Once these values are computed for each node in the two intervals the following are the alternatives we tried (from the simplest to more complicated):
\begin{itemize} 
\item  \textbf{Method 1:} Sort the nodes in each interval in  descending order by their cut size reduction.  Then simply do a   pairwise swap between entries at $i$th place in each interval provided that the swap results in a combined reduction.  This  method is very simple and fast, however, it may have  side effects within the interval (and also outside) since it does  not account for those. 

\item \textbf{Method 2:} This is a modified version of Method 1 with
   the addition that after each swap is performed we also update the
   cut reduction values of other nodes. This method does
   only one pass from top to bottom and stops when there is no further
   improvement.
  \item \textbf{Method 3:} In this method we iterate on Method 2 until we converge to a stable state. In this way we reach a local optimum between two intervals. The other improvement is that instead of pairing entries at $ith$ position we find the best pair in the second interval and swap with that. This is because the top entries in each interval can be neighbors of each other and therefore they may not be the best pair to swap.
 \end{itemize}
 
 Our extensive studies with these methods confirmed that Method 3 gives the best results in terms of quality at the expense of an acceptable additional cost. It runs fast in practice due to the relatively small sizes of the intervals because of the parallelism we achieve. We can also afford an additional cost in the swap method because each interval pair is handled by a separate Mapper task. 
\fi

%% file: postprocess.tex
\iffullversion
We have now an ordering of the vertices that presumably places similar
vertices close to each other. Instead of cutting the sequence at
equidistance positions to obtain a fully balanced partitioning, we
describe in this section several postprocessing techniques that allow
us to take advantage of the permissible imbalance and produce better
cuts.
\else
Here we several postprocessing techniques that allow
us to take advantage of the permissible imbalance and produce better
cuts.
\fi

First we sketch how the problem can be solved optimally if the number
of vertices is not too large. This approach is based on dynamic
programming and requires the entire graph to live in memory, which is
a barrier for using this technique even if the running time were
almost linear.\footnote{The procedure proposed here runs in time
  $O(n^3\log k)$ but the running time can be improved to almost
  linear.} However, this algorithm is still usable with the caveat
that one first needs to group together blocks of contiguous vertices
and contract them into supernodes, such that the graph of supernodes
fits in memory and is small enough for the algorithm to handle. The
fact that nearby vertices are similar to each other implies that this
sort of contraction should not hurt the subsequent optimization
significantly. Indeed trying two different block sizes (resulting in
$1000$ and $5000$ blocks) produced similar cuts.

%
The other two approaches work on the concept of ``windows'' and
perform only local optimizations within each. Let $\pi$ denote the
permutation of vertices at the start of the local optimization, and
let $\pi_i$ be the $i$-th vertex in the permutation for $1 \leq i\leq
n$. Given that we want to get $k$ parts of (almost) the same size, we
consider the ideal split points $q_j = \lfloor \frac{jn}{k} \rfloor$
for $1 \leq j < k$. Though there are only $k-1$ real split points, we
define two dummy split points at either end to make the notation
cleaner: $q_0 = 0$ and $q_k = n$. Then, the ideal partitions, in terms
of size, will be formed of $\pi(q_j+1 \rightarrow q_{j+1})$ 
for different $0 \leq j < k$. Given a permissible imbalance $\alpha >
0$, a window is defined around each (real) split point allowing for
$\alpha / 2$ imbalance on either side. More precisely, $W_j =
\pi(\pi_{q_j - \alpha'} \rightarrow \pi_{q_j + \alpha'})$
for $\alpha' =
\lceil \frac{\alpha n}{2k} \rceil$. Essentially we are free to shift
the boundaries of partitions within each window without violating the
balance requirements. We propose two methods to take advantage of this
opportunity. One finds the optimal boundaries within windows, while
the other also permutes the vertices in each window arbitrarily to
find a better solution.

%% file: dp.tex


The partitioning problem becomes more tractable if we fix an ordering
$\pi$ of vertices and insist on each partition consisting of
consecutive vertices (perhaps, with wrap-around).  
\iffullversion
Let us for now
assume that no wrap-around is permissible; i.e., each part corresponds
to a contiguous subset of vertices on $\pi$.
\else
Here we focus on the case with no wrap-around.
\fi

\iffullversion
In the dynamic programming (DP) framework, instead of solving the
given instance, we solve several instances (all based on the input and
closely related to one another), and do so in a carefully chosen
sequence such that the instances already solved make it easier to
quickly solve newer instances, culminating in the solution to the real
instance.
\fi

Let $A_{i,j,q}$ for $1\leq i, j \leq n, 1\leq q\leq k, j \geq i-1$
denote the smallest 
\iffullversion cut size achievable for the subgraph induced by
vertices $\pi(i  \rightarrow j)$ if exactly $q$ partitions thereof are desired.  
\else 
cut size, using exactly $q$ partitions,
achievable for the subgraph induced by vertices $\pi(i \rightarrow j)$.  
\fi 
Once all these entries are computed, $A_{1,n,k}$ yields the solution to the
original problem.  As is customary in dynamic programs, we only
discuss how to find the ``cost''---here, the cut size---of the optimal
solution; 
\iffullversion
finding the actual solution---here, the partition---can 
be achieved in a straight-forward manner by adding certain 
information to the DP table.
\else
augmenting the DP to find the actual solution---here, the
partition---is then standard.
\fi

We fill in the table $A_{i,j,q}$ in the order of increasing $q$.  For
the case of $q = 1$, we clearly have $A_{i,j,q} = 0$ if and only if
\iffullversion $$w(\pi(i \rightarrow j)) \leq (1+\alpha)
\frac{w(V)}{k},$$\else $w(\pi(i \rightarrow j)) \leq (1+\alpha) w(V) /
k$,\fi i.e., vertices $\pi(i \rightarrow j)$ \ifabstract fit in one
part.  \else can be placed in one part without violating the balanced
property.  \fi Otherwise, it is defined as infinity.

We use a recurisve formula to compute $A_{i,j,q}$ for $q>1$.
The formula depends on $A_{i',j',q'}$ entries where $q' < q$, hence
\iffullversion these entries have all been \fi already computed.
Let us introduce some notation before presenting the recurrence.
We define $C(i, j, k)$ as the total length of edges going from
$\pi(i \rightarrow j)$ to $\pi(j+1 \rightarrow k)$.
We now present the recursive formula for computing $A_{i,j,q}$ as
\begin{align}\label{eqn:recur1}
  A_{i,j,q} &= 
      \min_{i-1 \leq k \leq j}  \left[  A_{i,k,1} + A_{k+1,j,q-1} + C(i,k,j) \right].
\end{align}

\iffullversion Notice that the first term in the minimization is only
used to signify whether $\pi(i \rightarrow k)$ is a valid part in the
intended $\alpha$-balanced partition: its value is either zero or
infinity.  \fi

\begin{lemma}
  Equation~\eqref{eqn:recur1} is a valid recursion for $A_{i,j,q}$,
  which coupled with the initialization step given above yields
  a sound computation for all entries in the DP table.
\end{lemma}

\iffullversion
\begin{proof}
  The argument proceeds by mathematical induction. The initialization
  step is clearly sound as it is simply verifying feasiblity of single
  parts.

  For the inductive part, it suffices to show that the best solution
  with a part $\pi(i \rightarrow k)$ has a cut size $A_{i,k,1} +
  A_{k+1,j,q-1} + C(i,k,j)$.  Then, the first term in
  \eqref{eqn:recur1} accounts for the case when we use one empty part
  (i.e., we partition $\pi(i \rightarrow j)$ using only $q-1$ parts)
  and the minimization considers the cases when the first part
  consists of $\pi(i \rightarrow k)$.  To see the cost of thelatter is
  as in \eqref{eqn:recur1}, notice that any cut edge in the partition
  of $\pi(i \rightarrow j)$ is either completely inside $\pi(i
  \rightarrow k)$ (i.e., not contributing to the cut size), inside
  $\pi(k+1 \rightarrow j)$ (which is accounted for in $A_{k+1, j,
    q-1}$ or connects $\pi(i \rightarrow k)$ to $\pi(k+1 \rightarrow
  j)$ (in which case is taken into account by $C(i, k, j)$).
\end{proof}
\fi

The above procedure can be implemented to run in time $O(n^3k)$ and
consume space $O(n^2k)$.  \iffullversion Next we show how to improve
this further. \fi
The key idea \ifabstract for improving the running time \fi is to
avoid computing $A_{i, j, q}$ for all values $1\leq
q\leq k$.  Rather we focus on a limited subset of such $q$'s of size
$O(\log k)$.
\iffullversion

We start by rewriting the recurrence as
\else
Recurrence~\eqref{eqn:recur1} can be rewritten as
\fi
\begin{align}\label{eqn:recur2}
  A_{i,j,q} &= 
      \min_{i - 1 \leq k < j}  \left[  A_{i,k,\lfloor \frac{q}{2}\rfloor} + A_{k+1,j,\lceil \frac{q}{2}\rceil} + C(i,k,j) \right].
\end{align}
\iffullversion
Then, computing the desired value $A_{1,n,k}$ only requires the
computation of $A_{i,j,q}$ where 
 $q$ is either $\lfloor \frac{k}{2^l}
\rfloor$ or $\lceil \frac{k}{2^l} \rceil$ for some 
$l\geq 0$.  This
reduces the running time to $O(n^3\log k)$, and a bottom-up DP
computation needs no more than three different values for $q$, hence
the memory requirement is $O(n^2)$.
\else
Then, it suffices to compute $A_{i,j,q}$ when  
 $q \in \{\lfloor \frac{k}{2^l}\rfloor, \lceil \frac{k}{2^l} \rceil :
 l\geq 0\}$, reducing the running time to $O(n^3\log k)$, where a
 bottom-up DP computation needs no more than three different values
 for $q$, hence an $O(n^2)$ memory usage.
\fi



%% file: linear-opt.tex
\iffullversion
Recall the notion of ``windows'' defined at the beginning of
Section~\ref{sec:post}. 
\fi
We now focus on a window $W = W_j$. \ifabstract (See the definition at the
beginning of the section.)  \fi 
\iffullversion
This is small enough so 
\else
Note 
\fi
that all information on \ifabstract this \else the vertices in the \fi window
(including \ifabstract its vertices $V_W$ and \fi all their edges
\ifabstract $E_W$\fi) fit in memory (of one Mapper or Reducer).
The linear optimization postprocessing finds a new split point in each
window\iffullversion, so that the total weight of edges crossing it is
minimized.\else\ to minimize the total weight of edges crossing it.\fi

Let $B$ and $A$ denote the set of vertices appearing before and after
$W$, respectively, in the ordering $\pi$. The edges going from $B$ to
$A$ are irrelevant to \iffullversion the local optimization \else this
step \fi since \iffullversion those edges are necessarily cut no
matter what split point we choose\else they are cut regardless of the
split point\fi. The other edges have at least one endpoint in
$W$. Then, there is a simple \ifabstract $O(|V_W|\cdot|E_W|)$-time \fi
algorithm \iffullversion of running time $O(|V_W|\cdot|E_W|)$ \fi to
find the best split point in $W$\iffullversion\ where $V_W$ and $E_W$
denote the set of vertices and edges, respectively, corresponding to
$W$\fi: look at each candidate split point and go over all relevant
edges to determine the weight of the associated cut.

This can be done more efficiently to run in time $O(|V_W| + |E_W|)$,
too. Let $s_v$ for $v \in W$ be the cut value (ignoring the effect of
the edges between $B$ and $A$). We scan the candidate split points
from left to right, and compute $s_v - s_0$ where $s_0 =
s_{\pi_{q_j-\alpha'}}$ (i.e., the cut value for the first split point
in $W$).\footnote{Clearly, $s_0$ is the total weight of edges between
  $B$ and $W$ and itself can be computed in time $O(|V_W| + |E_W|)$
  (without hurting the overall runtime guarantee), however, the
  additive term $-s_0$ does not matter in comparing the different
  split point candidates.} For each vertex $v$ following $u$, start
with $s_u - s_0$, remove the edges between $v$ and the vertices to its
left (these already showed up in the cut value) and add up the edges
between $u$ and the vertices to its \iffullversion right; this gives
the cut value \else right to obtain \fi
$s_v - s_0$. 

Since the windows are relatively small compared to the entire graph,
and their corresponding subproblems can be solved independently, there
is a very efficient MapReduce implementation for window-based
postprocessing methods.


%% file: mincut-opt.tex
Once again we focus on a window $W = W_j$%
\iffullversion, and denote by $B$ and $A$
the set of vertices belonging to previous and following vertices.
Recall that the linear optimization of the previous section finds the
best split point in the window while respecting the ordering of the
vertices given in the permutation in $\pi$. The  minimum-cut
optimization, though, \fi
\ifabstract. Unlike the boundary optimization, the  minimum-cut
optimization is not bound to the ordering of vertices, and \fi
finds the best way to partition $W$ into two
pieces $W_L$ and $W_R$ so that the total weight of edges going from $B
\cup W_L$ to $A \cup W_R$ is minimized\ifabstract\ where $A, B$ were
defined above\fi.


\input{mincut-fig}

%% file: mincut-fig.tex



\begin{figure}
\tikzstyle{vert}=[fill=blue,circle,inner sep=.7mm]
\tikzstyle{vertr}=[draw=blue,fill=white,thick,circle,inner sep=.7mm]
\tikzstyle{wind}=[fill=yellow!60,rounded corners]
\centering
\subfigure[The ordering of vertices on the line prior to \mincut procedure for an instance with $k$ where the three windows are marked. The edges are not shown.]{
\begin{tikzpicture}
  \draw node (O) {};
  \foreach \i in {1,2,3}
   {
    \draw[wind] (O) ++(\i*20mm-10mm,2.5mm) rectangle ++(10mm, -5mm);
    \draw (O) ++(\i*20mm-10mm+5mm,5mm) node {$W_{\i}$};
   }
  \foreach \i in {1,...,56}
    \draw (O) ++(\i*1.25mm+.6mm+8.8mm-10mm, 0) node[vert,inner sep=.3mm] {};
\end{tikzpicture}
}
\ifabstract\vskip-1mm\fi
\subfigure[Minimum cut instance for window $W_2$. The diagram only shows edges with one endpoint in the window (and excludes any edges coming from beyond the neighboring windows since those are not affected by the choice of the cut). The source node, to which all edges going to the left are attached, and the sink node, to which all edges going to the right are attached, are not shown. ]{
\begin{tikzpicture}
  \draw node (O) {};
  \foreach \x in {1,2,3,4,5,6,7,8}
    \draw (O) +(\x*0.6,0) node[inner sep=0mm] (V\x) {};
\begin{scope}
  \clip (V1) +(-18mm,7mm) rectangle ([xshift=18mm,yshift=-16mm] V8);
  \draw (V1) +(-5,0) node (VL) {};
  \draw (V8) +(5,0) node (VR) {};
  \draw[wind] (V1) +(-7mm,6mm) rectangle ([xshift=7mm,yshift=-3mm] V8);
  \foreach \x/\ang/\t/\r in 
  {1/-160/L/3,1/-40/8/2,
   2/-20/R/4, 2/-30/R/4,
   3/-160/L/4,3/-140/L/3, 3/-40/R/3,
   4/-30/R/4,
   5/-20/R/3, 5/-30/R/3,
   6/-20/R/3, 7/-150/L/4}
     \draw (V\x.\ang) .. controls ++(\ang:\r{}cm) and ++(-180-\ang:\r{}cm) .. (V\t);
  \foreach \x in {1,2,3,4,5,6,7,8}
  \draw (V\x) node[vert,label=above:$\x$] {};
\end{scope}
\end{tikzpicture}
}
\ifabstract\vskip-1mm\fi
\subfigure[The solution from \linopt, which respects the ordering within the window and only optimizes for the split point. Solid and empty nodes are the two sides of the cut. The value of the cut is $4$.]{
\begin{tikzpicture}
  \draw node (O) {};
  \foreach \x in {1,2,3,4,5,6,7,8}
    \draw (O) +(\x*0.6,0) node[inner sep=0mm] (V\x) {};
\begin{scope}
  \clip (V1) +(-18mm,7mm) rectangle ([xshift=18mm,yshift=-16mm] V8);
  \draw (V1) +(-5,0) node (VL) {};
  \draw (V8) +(5,0) node (VR) {};
  \draw[wind] (V1) +(-7mm,6mm) rectangle ([xshift=7mm,yshift=-3mm] V8);
  \foreach \x/\ang/\t/\r in 
  {1/-160/L/3,1/-40/8/2,
   2/-20/R/4, 2/-30/R/4,
   3/-160/L/4,3/-140/L/3, 3/-40/R/3,
   4/-30/R/4,
   5/-20/R/3, 5/-30/R/3,
   6/-20/R/3, 7/-150/L/4}
     \draw (V\x.\ang) .. controls ++(\ang:\r{}cm) and ++(-180-\ang:\r{}cm) .. (V\t);
  \foreach \x/\y in {1/t,2/tr,3/tr,4/tr,5/tr,6/tr,7/tr,8/tr}
  \draw (V\x) node[ver\y,label=above:$\x$] {};
\end{scope}
\end{tikzpicture}
}
\ifabstract\vskip-1mm\fi
\subfigure[The solution from \mincut, which tampers with the ordering
within the window. \iffullversion Solid and empty nodes are the two sides of the cut.\fi The value of the cut is $1$.]{
\begin{tikzpicture}
  \draw node (O) {};
  \foreach \x/\y in {1/1,2/6,3/3,4/7,5/8,6/5,7/2,8/4}
    \draw (O) +(\y*0.6,0) node[inner sep=0mm] (V\x) {};
\begin{scope}
  \clip (V1) +(-18mm,7mm) rectangle ([xshift=18mm,yshift=-16mm] V5);
  \draw (V1) +(-5,0) node (VL) {};
  \draw (V8) +(5,0) node (VR) {};
  \draw[wind] (V1) +(-7mm,6mm) rectangle ([xshift=7mm,yshift=-3mm] V5);
  \foreach \x/\ang/\t/\r in 
  {1/-160/L/3,1/-40/8/1,
   2/-20/R/3, 2/-40/R/2,
   3/-160/L/4,3/-140/L/3, 3/-40/R/3,
   4/-30/R/2,
   5/-20/R/1, 5/-30/R/2,
   6/-20/R/3, 7/-150/L/4}
     \draw (V\x.\ang) .. controls ++(\ang:\r{}cm) and ++(-180-\ang:\r{}cm) .. (V\t);
  \foreach \x/\y in {1/t,2/tr,3/t,4/tr,5/tr,6/tr,7/t,8/t}
  \draw (V\x) node[ver\y,label=above:$\x$] {};
\end{scope}
\end{tikzpicture}
}
\caption{Illustration of \mincut postprocessing.}
\end{figure}
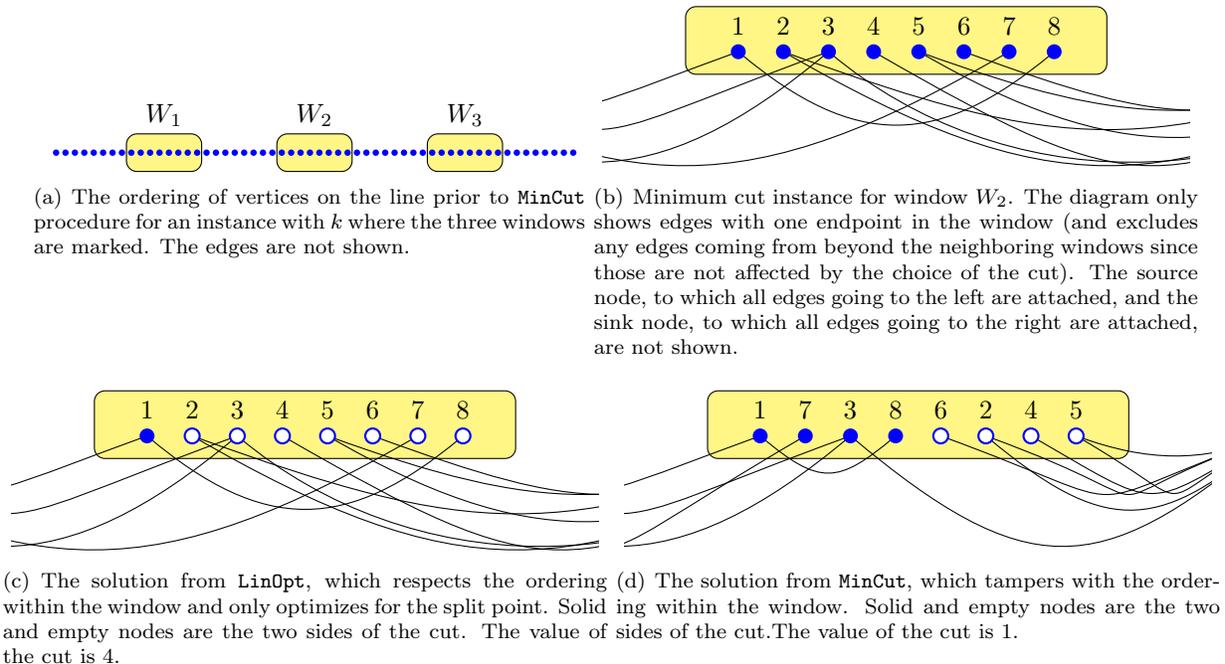

%% file: practice-main.tex
First we describe the different datasets that we use in our
experiments.
\iffullversion

\fi
Next we compare our results to previous work, and we finally compare
our different methods to each other: i.e., we demonstrate through
experiments how much value each ingredient of the algorithm adds to
the solution.


%% file: datasets.tex
We present our results on three datasets: \worldg, \twitterg and
\liveg (as well as publish our output on \friendg). As the names
suggest, the first one is a geographic dataset while the last three
are social graphs.
All are big graphs, representatives of maps and social networks, and
we test the quality and scalability of our algorithm on them.

\begin{description}
\item[\worldg] A subset of the entire world road network with hundreds
  of millions of vertices and over a billion edges.  (Due to
  near-planarity, the averagre degree is small.) \iffullversion Edges
  do not have weights\else The graph is unweighted\fi, but vertices
  have longitude/latitude \iffullversion information\else tags\fi.
  Our algorithms run smoothly on this graph in reasonable time using a
  small number of machines, demonstrating their scalability.
\item[\twitterg] The public graph of tweets\footnote{Other graphs
    in~\cite{fennel14} were either small or not public.}, with about
  41 million vertices (twitter accounts) and 2.4 billion (directed)
  edges (denoting followership)~\cite{Kwak2010-Twitter}.  This graph
  is unweighted, too. We run all our algorithms on the undirected
  underlying graph.
\item[\liveg] The undirected version of this public social graph
  (snapshot from 2006) has $4.8$ million vertices and $42.9$ million
  edges~\cite{wsdm-paper}.
\iffullversion
\item[\friendg] The undirected version of this public social graph
  (snapshot from 2006) has $65.6$ million vertices and $1.8$ billion
  edges.
\fi
\end{description}

%% file: ingreds.tex
\subsubsection{Initial ordering}
We have four methods to obtain an initial ordering for the geographic
graphs and two 
for non-geographic
graphs. Here we compare these methods to each other based on the value
of the cut obtained by chopping the resulting order into equal
contiguous pieces. In order to make this meaningful, we report all the
cut sizes as the fraction of cut edges to the total number of edges in
the graph.

\iffullversion
\begin{figure}
\centerline{\includegraphics[width=.6\columnwidth]{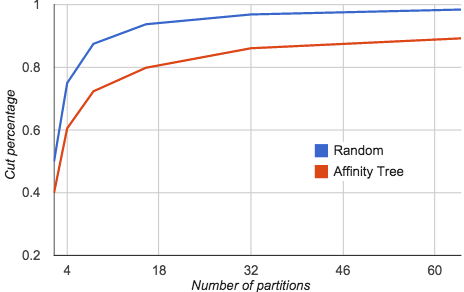}}
\caption{Comparison of the (fully balanced) cut size for \randinit
  and \affcomm on \twitterg.}\label{fig:initial-twitter}
\end{figure}
\fi

\ifabstract
\begin{figure}
\centerline{\includegraphics[width=.55\columnwidth]{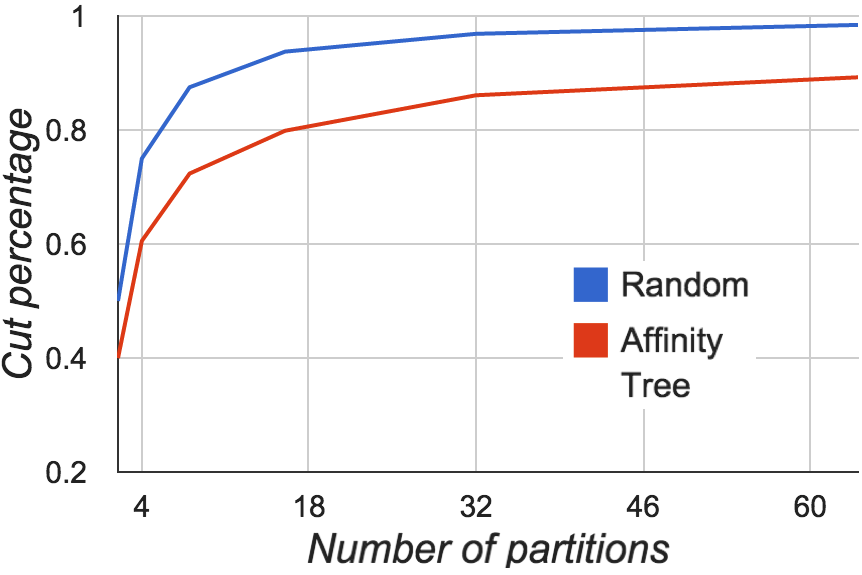}
\includegraphics[width=.55\columnwidth,height=3.1cm]{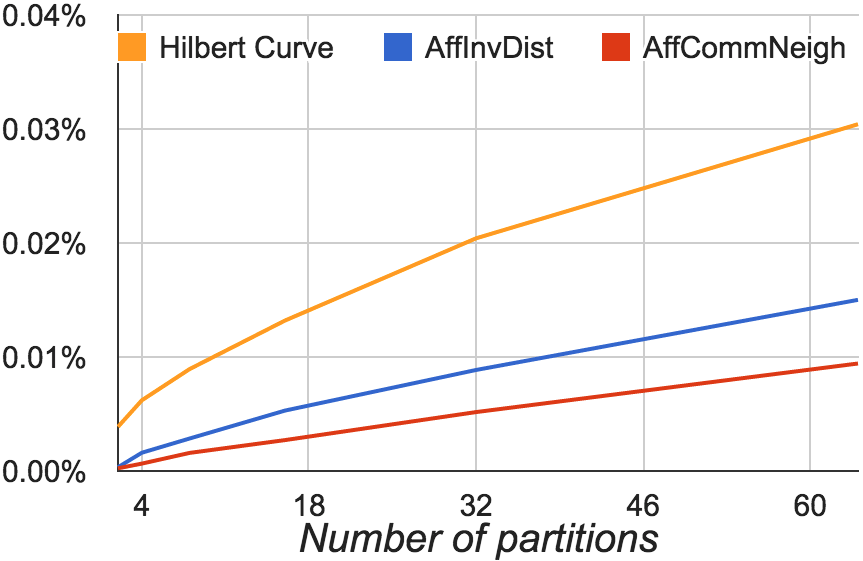}}
\caption{Comparison of the (fully balanced) cut size for \randinit
  and \affcomm on \twitterg (left) and \hilbert, \affdist
  and \affcomm on \worldg (right).}\label{fig:initial-twitter}\label{fig:initial-world}
\end{figure}
\fi

Figure~\ref{fig:initial-twitter} compares the results of two methods:
\randinit produces a random permutation of the vertices, hence, 
as we observed prevously,
gets a cut size of approximately $1-\frac{1}{k}$; \affcomm (using the
\affinity with the similarity oracle based on the number of common
neighbors) clearly produces better solutions with improvements ranging
from 20\% to 10\% (more improvement for smaller number of parts).  The
\affcomm tree in this case was pretty shallow (much lower than
\iffullversion the theoretical \fi $\log n$ \iffullversion upper
bound\fi) with only four levels. Whenever we report numbers for
different number of partitions in one chart, $k = 2, 4, 8, 16, 32, 64$
are used for experiments.

Figure~\ref{fig:initial-world} \ifabstract also \fi compares the results of three methods,
each producing a cut of size at most $3\%$ and significantly improving
upon the $1-\frac{1}{k}$ result of \randinit: Compared to \hilbert,
the two other methods \affdist (using the inverse of geographic
distance as the similarity oracle) and \affcomm, respectively, obtain
$90\%$ to $50\%$ and $95\%$ to $70\%$ improvement; once again the
easier instances are those with smaller $k$.  The corresponding trees,
with 13 and 11 levels, respectively, are not as shallow as the one for
\twitterg.
The quality of the cuts correlates with the runtime
complexity of the algorithms: \randinit (fastest, but worst),
\hilbert, \affdist, and \affcomm (best, but slowest).  Note that
although the implementation of \hilbert seems more complicated, it is
significantly faster in practice.

\iffullversion
\begin{figure}
\centerline{\includegraphics[width=.6\columnwidth]{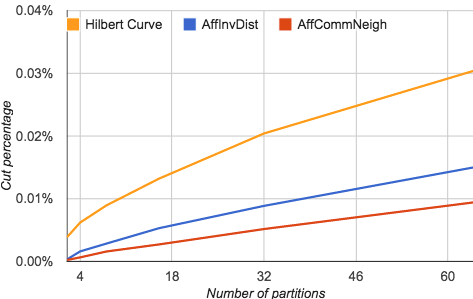}}
\caption{Comparison of the (fully balanced) cut size for \hilbert, \affdist
  and \affcomm on \worldg.}\label{fig:initial-world}
\end{figure}
\fi

For the following experiments, we focus on \affcomm since it
consistently \iffullversion produces \else gives \fi better results for different values of $k$
\iffullversion as well as both for \twitterg and \worldg\else (both
for \twitterg and \worldg)\fi.  Specially in diagram legends, this
initialization step may be abbreviated as \verb|Aff|.

\subsubsection{Improvements and imbalance}

\iffullversion
\begin{figure}
\centerline{\includegraphics[width=.6\textwidth]{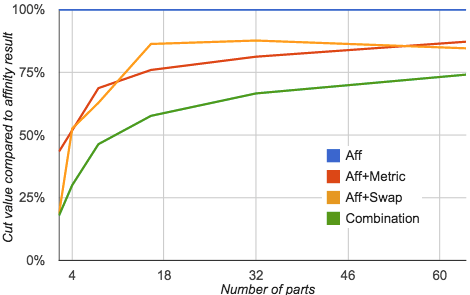}}
\caption{Comparison of the (fully balanced) cut size for three
  semilocal operations on \twitterg. The results are compared to
  \affcomm, set as 100\%.}\label{fig:imp-twitter}
\end{figure}
\fi

\ifabstract
\begin{figure}
\centerline{\includegraphics[width=.57\columnwidth]{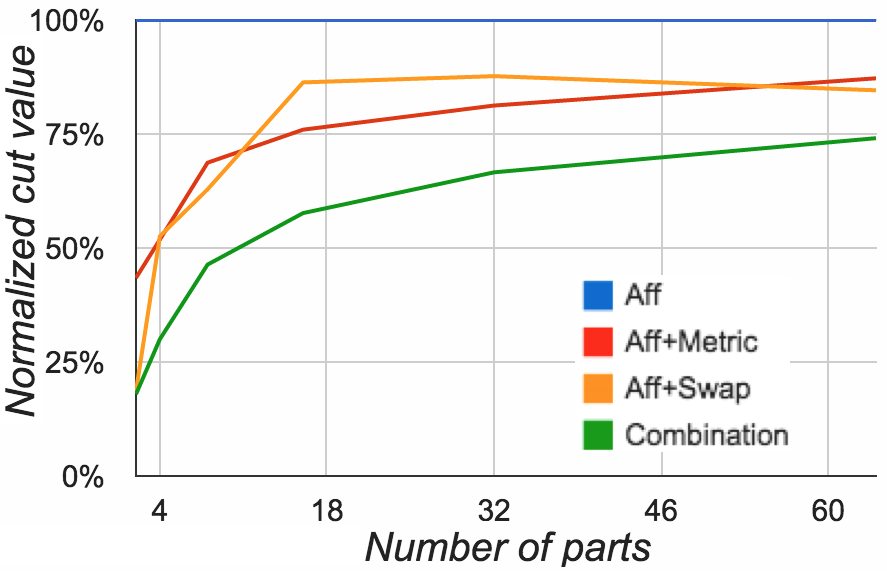}
\includegraphics[width=.53\columnwidth,height=3cm]{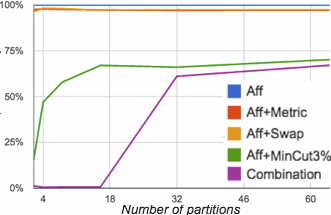}}
\caption{Comparison of the (fully balanced) cut size for some
  semilocal operations on \twitterg (left) and fully balanced and
  5\%-imbalance cuts for \worldg (right). The results are compared to
  \affcomm, set as 100\%.}\label{fig:imp-twitter}\label{fig:all-world}
\end{figure}
\fi

The performance of the semilocal improvement methods, \affmetric and
\affswap, as well as our main algorithm, \champion, on \twitterg is
reported in Figure~\ref{fig:imp-twitter}.  The diagram shows the
percentage of improvement of each over the baseline \affcomm for
different values of $k = 2, \iffullversion 4, \dots, 32, \else \dots \fi 64$.  Except for $k=16, 32$,
\affswap outperforms \affmetric; their performance varies from $7\%$
to $75\%$, with best performance for smaller $k$.  Combining the two
methods with the imbalance-inducing techniques to get \champion yields
results significantly better than every single ingredient, with the
relative improvement sometimes as large as 50\%.  The
final results have very little imbalance, and indeed cuts of comparable
quality can be obtained without any imbalance.


\iffullversion
\begin{figure}
\centerline{\includegraphics[width=.6\textwidth]{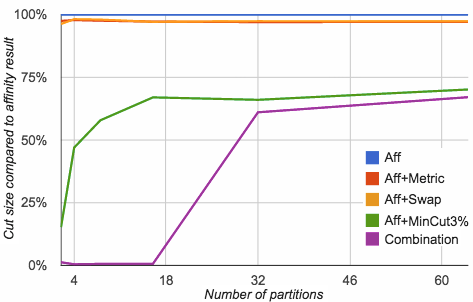}}
\caption{Cut value obtained by semilocal optimization methods on
  \worldg as compared to their baseline, \affcomm.}\label{fig:all-world}
\end{figure}
\fi

Figure~\ref{fig:all-world} also sets the results of various postprocessing
methods against one another and depicts their improvements compared to
the baseline of \affcomm when run on \worldg. The first two consists
of semilocal optimization methods, \affmetric and \affswap, which also
appeared in the discussion for \twitterg. These do not yield
significant improvement over their \affcomm starting point: the
results compared to the starting point (baseline) ranges from 95\% to
98\%.  The other two algorithms use imbalance-inducing ideas, and
yield significantly better results.  Our main algorithm,
\iffullversion which we call \fi
\champion, clearly outperforms all the others specially for small
values of $k$.

All the experiments assume $3\%$ imbalance in partition sizes---each
partition may have size within range $(1\pm 0.03) \frac{n}{k}$.
However, only \affmincut and \champion use this to improve the quality
of the partition.

\subsubsection{Convergence analysis}

\ifabstract
\begin{figure*}
\centering
\subfigure {\includegraphics[width=.50\columnwidth]{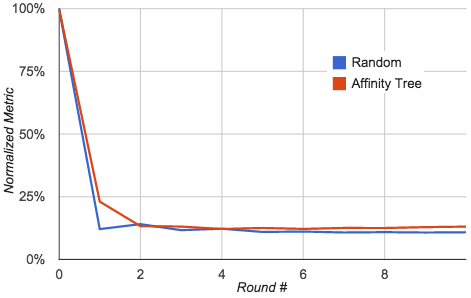}}
\subfigure {\includegraphics[width=.50\columnwidth]{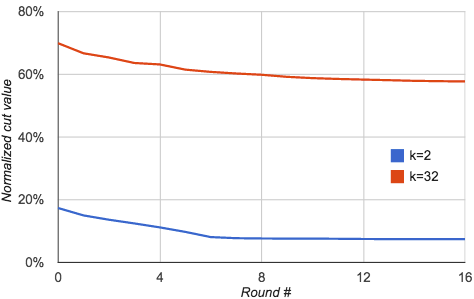}}
\subfigure{\includegraphics[width=.50\columnwidth]{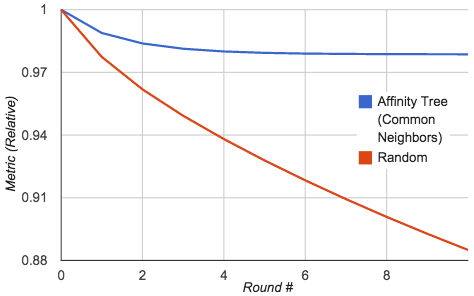}}
\subfigure{\includegraphics[width=.50\columnwidth]{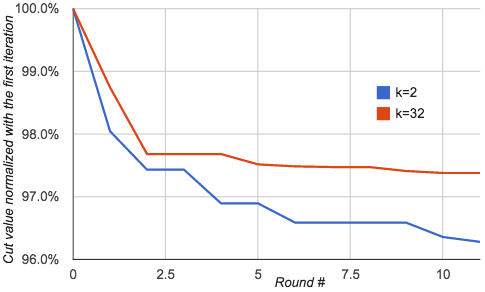}}
\caption{Convergence rate of \affmetric (first) and \affswap (second) on \twitterg, and then convergence rate of \affmetric (third) and \affswap (fourth) on \worldg.}\label{fig:world:converge}\label{fig:twitter:converge}
\end{figure*}
\fi

\iffullversion
\begin{figure}
\centerline{\includegraphics[width=.6\textwidth]{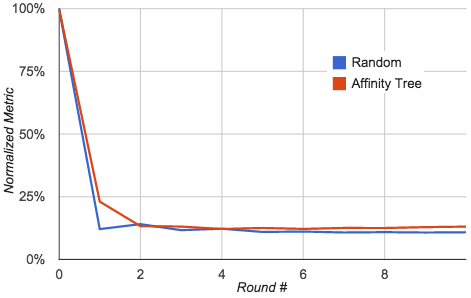}}
\centerline{\includegraphics[width=.6\textwidth]{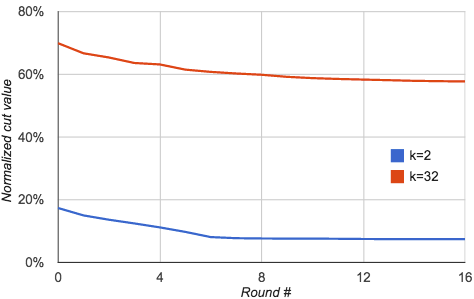}}
\caption{Convergence rate of \affmetric (top) and \affswap (bottom) on \twitterg.}\label{fig:twitter:converge}
\end{figure}
\fi
Next we consider the convergence rate of the two semilocal improvement
methods on two graphs. For \twitterg, the rate of convergence for
metric optimization is the same whether we start with \affcomm or
\randinit. (See Figure~\ref{fig:twitter:converge}.) The convergence
happens essentially in three or four rounds.

The convergence for the rank swap methods happens in 10-15 steps for
\twitterg and in 5-10 steps for \worldg; more steps required for larger
values of $k$.


The convergence for the metric optimization on \twitterg happens in 3-4
steps if the starting point is \affcomm.  However, if we start with a
random ordering, it may take up to 20 steps for the metric to converge.

\iffullversion
\begin{figure}
\centerline{\includegraphics[width=.6\textwidth]{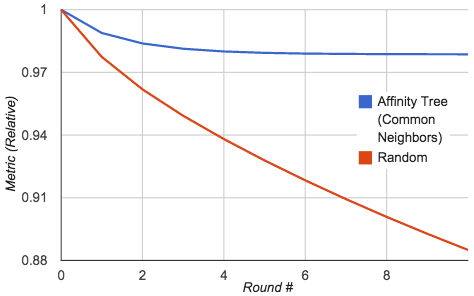}}
\centerline{\includegraphics[width=.6\textwidth]{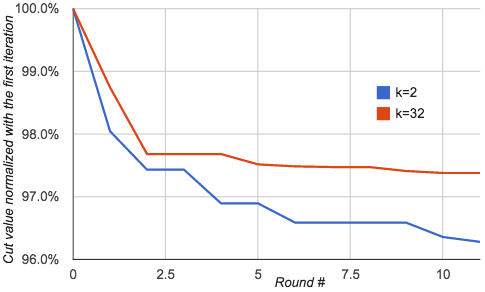}}
\caption{Convergence rate of \affmetric (top) and \affswap (bottom) on \worldg.}\label{fig:world:converge}
\end{figure}
\fi

%% file: comparison-facebook.tex
The most relevant \iffullversion among the \fi previous work are the scalable label
propagation-based algorithms of Ugander and Backstrom~\cite{wsdm-paper}
  and Martella et al.~\cite{Spinner14}.  The former reports
  \iffullversion their \fi results on
an internal graph (for Facebook) and a public social graph, \liveg,
while the latter reports results on \liveg and \twitterg.
\iffullversion The cut sizes in the table for $k=40, 60, 80$ for
\cite{wsdm-paper} and all the numbers for \cite{Spinner14} are
approximates taken from the graph provided in \cite{wsdm-paper} as
they do not report the exact values.\fi 
The \iffullversion numbers in parentheses show the maximum imbalance
in partition sizes, \else
maximum partition imbalance is given in parentheses, 
\fi
and the cut sizes are
reported as fractions of total number of edges in the graph.

\vspace{1mm}
\centerline{
\begin{tabular}{|c|c|c|c|c|}
  \hline
 $k$ & $\substack{\mbox{\cite{wsdm-paper}}\\[-0.5mm](5\%)}$ &
 $\substack{\mbox{\texttt{Spinner}}\\(5\%)}$ & $\substack{\mbox{\texttt{Aff}}\\(0\%)}$ & $\substack{\mbox{\champion}\\(0\%)}$
 \\\hline
 20  &  37\%  &  38\% &  35.71\%  &  \textbf{27.50\%}  
 \\
 40  &  43\%   &  40\% & 40.83\%  &  \textbf{33.71\%}  
 \\
 60  &  46\%   &  43\% & 43.03\%  &  \textbf{36.65\%}  
 \\
 80  &  47.5\%   &  44\% & 43.27\%  &  \textbf{38.65\%}  
 \\
100  &  49\%   &  46\% & 45.05\%  &  \textbf{41.53\%}  
\\  \hline
\end{tabular}}
\iffullversion
\vspace{1mm}
\fi

Even our initial linear embedding (obtained by constructing the
hierarchical clustering based on a weighted version of the input,
where edge weights denote the number of common neighbors between two
vertices) is consistently better than the best previous result.  Our
algorithm with the post-processing obtains $15\%$ to $25\%$
improvement over the previous results (better for smaller $k$), and it
only requires a couple of post-processing rounds to converge.  We take
note that the results of \cite{wsdm-paper} allow $5\%$ imbalance
whereas our results produce almost perfectly balanced partitions.

%% file: comparison-fennel.tex
\twitterg is another public graph, for which the results of
state-of-the-art minimum-cut partitioning is available.  We report and
compare the results of our main algorithm, \champion, for this graph
to the best previous methods.\iffullversion\footnote{Numbers for \cite{SK12}
       are quoted from \cite{Spinner14}.}\fi

\iffullversion
\vspace{1mm}
\fi
\centerline{
\begin{tabular}{|c|c|c|c|c|c|}
  \hline
 $k$ & $\substack{\mbox{\cite{SK12}}\\(4-15\%)}$ &
 $\substack{\mbox{\tt Spinner}\\(5\%)}$ & $\substack{\mbox{\tt
     FENNEL}\\(10\%)}$ & $\substack{\mbox{\tt METIS}\\(3\%)}$ & $\substack{\mbox{\small\champion}\\(3\%)}$
 \\\hline
 2  & 34\% &  15\% & \textbf{6.8\%}  &  11.98\%  &  7.43\%
 \\
 4  & 55\% &  31\% & 29\%   &  24.39\%  &  \textbf{18.16\%}
 \\
 8  & 66\% &  49\% & 48\%   &  35.96\%  &  \textbf{33.55\%}
 \\
 16  & 76\% & 61\% & 59\%   &  N/A  &  \textbf{46.18\%}
 \\
 32  & 80\% & 69\% & 67\%   &  N/A  &  \textbf{57.67\%}
\\  \hline
\end{tabular}}
\iffullversion
\vspace{1mm}
\fi



Algorithms developed in \cite{SK12,fennel14} are suitable for the
streaming model, but one can implement variants of those algorithms in
a distributed manner. Our implementation of a natural distributed
version of the FENNEL algorithm does not achieve the same results as
those reported for the streaming implementation reported in
\cite{fennel14}. However, we compare our algorithm directly to the
numbers reported on \twitterg by FENNEL~\cite{fennel14}.

%% file: comparison-friendster.tex
Finally we report the cut sizes for \iffullversion another public
graph, \fi\friendg, so others can compare their results with it.  The
running time of our algorithms are not affected with large $k$; the
running time difference for $k=2$ and tens of thousands is less than
$1\%$, well within the noise associated with the distributed system.
In fact, construction of the initial ordering is independent of $k$,
and the post-processing steps may only take advantage of the increased
parallelism possible for large $k$.

\vspace{1mm}
\centerline{
\begin{tabular}{|c|ccc|}
  \hline
 $k$ & 2 & 10 & $10^2$ 
 \\\hline
Cut size  &  11.9\%   &  41.4\%  &  59.8\% 
\\  \hline
\end{tabular}}
\iffullversion
\vspace{1mm}
\fi

We also make our partitions for \twitterg, \liveg and \friendg
publicly available~\cite{public-output}.

%% file: scale.tex
We noted above that the choice of $k$ does not affect the running time
of the algorithm significantly: the running time for two and
tens of thousands of partitions differed less than $1\%$ for \friendg.

As another measure of its scalability, we run the algorithm with $k=2$ on a
series of random graphs (that are similar in nature) of varying sizes.
In particular, we use RMAT graphs with parameter 20, 22, 24, 26 and
28, whose node and edge count is given in the table below.  In
addition, the last column gives the normalized running time of our
algorithm on these graphs.  Note that the size of the graph almost
quadruples from one graph to the next.

\vspace{1mm}
\centerline{
\begin{tabular}{|l|c|c|c|c|}
  \hline
graph & $|V|$ & $|E|$ & max degree & running time
 \\\hline
RMAT 20 & 650K  & 31M  &  65K &  100\%
 \\
RMAT 22 & 2.4M  &  130M   &  160K &  110\%
 \\
RMAT 24 & 8.9M  &  525M   & 400K &  133\%
 \\
RMAT 26 & 32.8M  & 2.1B   &  1M &  160\%
 \\
RMAT 28 & 120M  &  8.5B   & 2.5M  &  402\%
\\  \hline
\end{tabular}}
\iffullversion\vspace{1mm}\fi


%% file: apps.tex
As discussed in Section~\ref{sec:intro}, we apply our results to the
Google Maps Driving Directions application.  Note that the evaluation
metric for balanced partitioning of the world graph is the expected
percentage of cross-shard queries over the total number of
queries. More specifically, this objective function can be captured as
follows: we first estimate the number of times each edge of graph may
be part of the shortest path (or the driving direction) from the
source to destination across all the query traffic. This produces
an edge-weighted graph in which the weight of each edge is
proportional to the number of times we expect this edge appearing in a
driving direction, and our goal is to partition the graph into a small
number of pieces and minimize the total weight of the edges cut. We
can estimate each weight based on historical data and use it as a
proxy for the number of times the edge appears on a driving direction
on the real data. This objective is aligned with minimizing the
weighted cut, and we can apply our algorithms to solve this problem.

Before deciding about our live experiments, we realized it might
be suitable to use an imbalance factor in dividing the graph, and as a
result, we first examined the best imbalance factor for our
cut-optimization technique. 
\iffullversion
The result of this study is summarized in
Figure~\ref{fig:pathfinder}. 
\fi 
In particular, we observe that we can
reduce cross-shard queries by $21\%$ when increasing the imbalance
factor from $0\%$ to $10\%$.

The two methods that we examined via live experiments were (i) a
baseline approach based on the Hilbert-curve embedding, and (ii) one
method based on applying our cut-optimization postprocessing
techniques. Note that both these algorithms first compute an embedding
of nodes into a line, which results in a much simpler system to
identify the corresponding shards for the source and destination of
each query at serving time. Finally, by running live experiments on
the real traffic, we observe the number of multi-sharded queries from
our cut-optimization techniques is almost 40$\%$ less than the number of
multi-sharded queries compared to the na\"\i{}ve Hilbert embedding technique.

\iffullversion
 \begin{figure}
\ifabstract
\centerline{\includegraphics[width=.6\columnwidth]{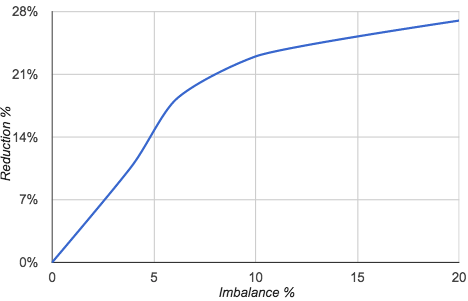}}
\else
\centerline{\includegraphics[width=.6\columnwidth]{img/pathfinder-imbalance.png}}
\fi
\caption{The percentage of improvement in cut value as we increase the allowable imbalance factor.}\label{fig:pathfinder}
\end{figure}
\fi

%% file: conclusion.tex
We develop a scalable algorithm for balanced partitioning of
geographic and general graphs.  The algorithm is based on embedding
the vertices onto a line before chopping the sequence of vertices into
almost equal pieces.  The initial ordering obtained from hierarchical
agglomerative clustering using the number of common neighbors as
similarity measure outperforms the Hilbert cover-based partitioning
even for geographic graphs.  While the most effective post\-processing
techniques are random-swap local improvements and minimum cut-based
boundary optimization, iterative use of different techniques until
convergence gives best overall results.  Our algorithms improves
upon the previously developed distributed balanced partitioning
algorithms ~\cite{wsdm-paper,fennel14,Spinner14}.  In addition, live
experiments for the Google Maps Driving Direction shows significant
advantage of our algorithm over a standard Hilbert curve-based
partitioning.

{\bf \noindent Acknowledgements.} The authors wish to thank Aaron
Archer and Raimondas Kiveris for fruitful discussions.